\documentclass[aps,prl,twocolumn]{revtex4}

\usepackage{amsmath, amssymb}
\usepackage{graphicx}

\renewcommand{\section}[1]{{\par\it #1.---}\ignorespaces}

\begin{document}
\title{Solid state Stern-Gerlach spin-splitter for magnetic field sensoring, spintronics, and quantum computing}

\author{Kristofer Bj\"{o}rnson}
\affiliation{Department of Physics and Astronomy, Uppsala University, Box 516, S-751 20 Uppsala, Sweden}
\author{Annica M. Black-Schaffer}
\affiliation{Department of Physics and Astronomy, Uppsala University, Box 516, S-751 20 Uppsala, Sweden}
\date{\today}

\begin{abstract}
We show that the edge of a two-dimensional topological insulator can be used to construct a solid state Stern-Gerlach spin-splitter.
By threading such a Stern-Gerlach apparatus with a magnetic flux, Ahranov-Bohm like interference effects are introduced.
Using ferromagnetic leads, the setup can be used to both measure magnetic flux and as a spintronics switch.
With normal metallic leads a switchable spintronics NOT-gate can be implemented.
Furthermore, we show that a sequence of such devices can be used to construct a single-qubit $SU(2)$-gate, one of the two gates required for a universal quantum computer.
The field sensitivity, or switching field, $b$ is related to the device characteristic size $r$ through $b = \frac{\hbar}{qr^2}$, with $q$ the unit of electric charge.
\end{abstract}

\maketitle
Two famous examples of the fundamental difference between quantum mechanical and classical particles are provided through the Stern-Gerlach (SG) experiment \cite{ModernQuantumMechanics} and the Aharanov-Bohm (AB) effect \cite{PhysRev.115.485}.
The SG experiment demonstrates the peculiar behavior of the quantum mechanical spin, teaching us that for any chosen axis the spin can be pointing either up or down. Even more unintuitive, the spin can also be in a superposition of these two states, and thereby split in a SG apparatus to travel along different paths \cite{ModernQuantumMechanics}.
The AB effect, on the other hand, shows that the introduction of a magnetic vector potential has important effects on the phase of the wave function.
This is not merely a mathematical formality, but has measurable consequences in interference measurements.
When a particle travels along two different paths that enclose a magnetic flux, it picks up different phases along the two paths, even though the paths do not pass through the magnetic flux \cite{PhysRev.115.485}.

A topological insulator is a material with insulating bulk, but with topologically protected helical edge states.
Here we show that it is possible to construct a solid state SG apparatus, or spin-splitter, using the edge states in a two-dimensional topological insulator (2D TI) \cite{PhysRevLett.95.226801, PhysRevLett.95.146802, PhysRevLett.96.106802, Science.314.1692, Science.314.1757, Science.318.766, RevModPhys.82.3045, RevModPhys.83.1057, PhysRevLett.107.076802}.
The device consists of a small hole drilled in the 2D TI, contacted by two leads.
By threading a magnetic flux through the hole, an AB-like effect gives rise to important interference effects, which allows for precise manipulation of spin currents.
While the ordinary AB effect arises because of interference in a single complex number, the effects achieved here relies on modifying the relative phase between the up and down components of the spin. Thus, the effects described here can be classified as $SU(2)$-AB effects, while the ordinary situation corresponds to an $U(1)$-AB effect.

While, the AB effect has recently attracted some attention in 3D TI \cite{PhysRevLett.105.206601, PhysRevLett.105.156803, NatureMat.9.225, NatureComms.6.7634, arXiv.1410.5823}, we here outline several concrete and different applications of the $SU(2)$-AB effect in 2D TI.
More specifically, we find that if using ferromagnetic leads, the device can be used for sensitive measurements of magnetic field strengths. The same setup can also be used to implement a spintronic switch.
Instead using normal metallic leads, a switchable NOT spintronics gate can easily be constructed.
Finally, we also demonstrate how a sequential setup of normal lead solid state SG spin-splitters can be used to construct a single-qubit $SU(2)$-gate, one of two gates required to construct a universal quantum computer \cite{PhysRevA.52.5.52}.

\section{Setup}
\begin{figure}
	\begin{center}
		\includegraphics[width=0.47\textwidth]{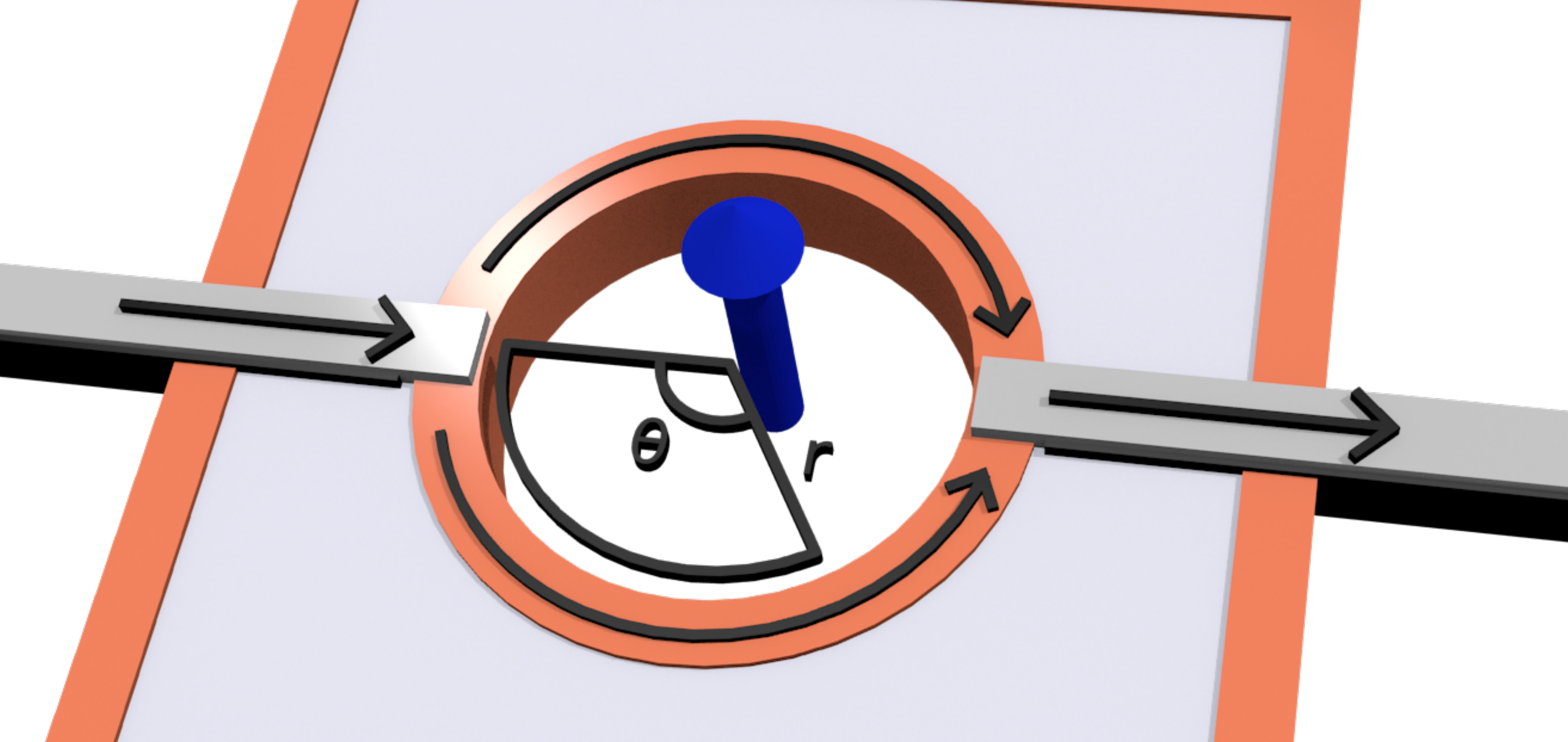}
	\end{center}
	\caption{A hole drilled in a 2D TI creates two edge channels (orange).
		Leads (grey) are attached on each side of the hole, and a bias voltage is applied across the circuit. The transport properties of the device can be altered by threading a magnetic flux (blue arrow) through the hole, as well as by choosing either ferromagnetic or normal leads. The circular shape is not essential, but is used to simplify calculations.}
	\label{Figure:Setup}
\end{figure}

Consider the setup in Fig.~\ref{Figure:Setup}.
The circular channel around the hole is an edge of a 2D TI and therefore hosts helical edge states.
We assume for simplicity that the spin-polarization axis is perpendicular to the plane of the TI. The Hamiltonian describing the two counter-propagating edge channels is then given by
\begin{align}
	H_{\uparrow} = -\frac{i\hbar}{r}\partial_{\theta},\nonumber\\
	H_{\downarrow} = \frac{i\hbar}{r}\partial_{\theta}\nonumber,
\end{align}
where arrows indicate the spin direction.
In the ground state no net current is carried from one side to the other, since the system is symmetric under a rotation of $\pi$ around the the $z$-axis orthogonal to the TI.
However, if a voltage is applied across the circuit, electrons can start to flow from one side to the other, say from the left to the right.
This current will be proportional to the transfer matrix of the states that are occupied at the left side, but unoccupied on the right. We therefore begin by calculating this transfer matrix.

When considering processes that transfers electrons from the left to the right, we can, because of the helicity of the edge states, restrict ourselves to up-spins along the upper edge, and down-spins along the lower edge.
Further, we introduce the coordinate $x_1 = r(2\pi - \theta)$ and $x_2 = r\theta$ along the upper and lower edges, respectively.
The eigenvalue equations along the two edges are then
\begin{align}
	H_{\uparrow}\Psi_{\uparrow p} =& i\hbar\partial_{x_1}\Psi_{\uparrow p} = E\Psi_{\uparrow p},\nonumber\\
	H_{\downarrow}\Psi_{\downarrow p} =& i\hbar\partial_{x_2}\Psi_{\downarrow p} = E\Psi_{\downarrow p},\nonumber
\end{align}
and the corresponding eigenstates can be written as
\begin{align}
	|\Psi_{\uparrow p}\rangle =& \left[\begin{array}{c}
					1\\
					0
				\end{array}\right]e^{-ipx_1/\hbar},\nonumber\\
	|\Psi_{\downarrow p}\rangle =& \left[\begin{array}{c}
					0\\
					1
				\end{array}\right]e^{-ipx_2/\hbar}.\nonumber
\end{align}

We now thread a magnetic flux of magnetic field strength $B$ through the hole.
To describe this we choose the vector potential $A = \frac{B}{2}r\hat{\theta}$, which translates into $-\frac{B}{2}r\hat{x}_1$ and $\frac{B}{2}r\hat{x}_2$ in the new $(x_1, x_2)$-coordinates.
The addition of this vector potential acts on the phase of the eigenstates according to
\begin{align}
	|\Psi_{\uparrow p}\rangle =& \left[\begin{array}{c}
					1\\
					0
				\end{array}\right]e^{-i(p + qBr/2)x_1/\hbar},\nonumber\\
	|\Psi_{\downarrow p}\rangle =& \left[\begin{array}{c}
					0\\
					1
				\end{array}\right]e^{-i(p - qBr/2)x_2/\hbar},\nonumber
\end{align}
where $q$ is the unit of electric charge.
It is therefore clear that the transfer matrix, which describes the transport of spins from the left side, $x_1 = x_2 = 0$, to the right side, $x_1 = x_2 = r\pi$, is given by
\begin{align}
	\label{Equation:Transfer_matrix_O}
	O = e^{-ipr\pi/\hbar}\left[\begin{array}{cc}
			e^{-iqBr^2\pi/2\hbar}	& 0\\
			0						& e^{iqBr^2\pi/2\hbar}
		\end{array}\right].
\end{align}
We here note that under a gauge transformation $A \rightarrow A + A'$, where $A'$ satisfies $\int_{0}^{2\pi}A'(\theta)d\theta = 0$, the transfer matrix transforms as $O \rightarrow Oe^{i\frac{q}{\hbar}\int_{0}^{\pi}A'(\theta)d\theta}$.
We have confirmed that this additional phase drops out of all physical quantities below, proving the gauge invariance of our results, and we can therefore set $A' = 0$.
Similarly, the overall phase in the above equation will drop out of all physical quantities.
This also justifies us in not having specified the chemical potential.
Because, as long as the spectrum is described by the same edge Hamiltonian, the only role of the chemical potential is to determine around what momentum $p_f$ the relevant excitations are located.

\section{Transfer between lead and edge channels}
The total transfer matrix for the system will not only depend on the transfer matrix that describes the motion around the hole, but also on the matrices that describe the transfer processes between the leads and the circular edge.
We will here assume that this process preserves phase coherence between the states in the leads and the TI edge states, and that it is described by a single tunneling parameter $t$, which we for now set to $t = 1$ to indicate perfect transmission between lead and edge.
That is, the transmission is described by the identity matrix, and therefore contributes trivially to the total transfer matrix.
However, we will in what follows be interested in tilting the TI by an angle $\varphi$ relative to the quantization axis of the leads. It is therefore necessary to also let the total transfer matrix encode a change of basis between the leads and the TI.
For this purpose we define two sets of coordinate axes, the laboratory axes $x,y,z$, and the TI axes $x',y',z'$.
We choose to describe the electrons in the leads with the coordinates in the laboratory frame, while the edge states in the TI are described by the primed coordinates.
It is clear that Eq.~\eqref{Equation:Transfer_matrix_O} refers to the transfer of states in the primed basis.
In particular, we choose the $x,x'$-axes along the direction of motion of the electrons through the circuit, while the $z,z'$-axes are chosen such that they coincide when $\varphi = 0$ and $z'$ is always perpendicular to the TI.
Explicitly, the $x,y,z$- and $x',y',z'$-coordinates are related through
\begin{align}
	x'	& =  x,\nonumber\\
	y'	& =  y\cos(\varphi) - z\sin(\varphi),\nonumber\\
	z'	& =  y\sin(\varphi) + z\cos(\varphi).\nonumber
\end{align}
The change of basis from the $x,y,z$-basis to the $x',y',z'$-basis for the spins is then given by
\begin{align}
	\label{Equation:Transfer_matrix_LR}
	L \equiv R^{\dagger}	 \equiv \left[\begin{array}{cc}
		\cos(\frac{\varphi}{2})		&	-i\sin(\frac{\varphi}{2})\\
		\sin(\frac{\varphi}{2})		&	i\cos(\frac{\varphi}{2})
	\end{array}\right].
\end{align}
We have here used $L$ and $R$ to denote the transformations from the unprimed to the primed coordinates, and the primed to the unprimed coordinates, respectively.
The symbols $L$ and $R$ are chosen since they are applied at the left and right end of the system, respectively.
With these definitions we are now ready to write down the complete transfer matrix for the system
\begin{align}
	T(B,r, \varphi) = ROL.\nonumber
\end{align}
Here we have made explicit the dependence of $T$ on the parameters $B$ and $r$ from Eq.~\eqref{Equation:Transfer_matrix_O}, and on $\varphi$ from Eq.~\eqref{Equation:Transfer_matrix_LR}.
The main advantage of introducing the $L$ and $R$ matrices is that they allow us to work in the laboratory frame alone.
To calculate the probability that an incoming spin $\sigma$ in the left lead is transferred to a spin $\lambda$ in the right lead, we now simply need to calculate the square of the corresponding matrix element
\begin{align}
	|T_{\lambda\sigma}|^2 = |\langle\lambda|T(B,r,\varphi)|\sigma\rangle|^2.\nonumber
\end{align}

\section{Measuring magnetic flux}
As a first example of a concrete application, we consider a system with fully spin-polarized ferromagnetic leads only containing electrons with spin-up.
Further, the SG spin-splitter is assumed to be oriented at an angle $\varphi = \pi/2$, which forces the incoming spins to split equally into both channels.
Because the leads only conduct spin-up electrons, the only relevant matrix element for the scattering matrix is
\begin{align}
	T_{\uparrow\uparrow} =& \langle\uparrow|T(B,r,\frac{\pi}{2})|\uparrow\rangle = \cos(\frac{qBr^2\pi}{2\hbar})e^{-ipr\pi/\hbar}.\nonumber
\end{align}
The conductance is therefore given by
\begin{align}
	\label{Equation:Conductivity}
	G =& \frac{e^2}{\hbar}|T_{\uparrow\uparrow}|^2 = \frac{e^2}{\hbar}\cos^2\left(\frac{qBr^2\pi}{2\hbar}\right).
\end{align}
It is clear that the prominent dependence of the current on the magnetic flux $Br^2\pi$ makes this setup ideal for measuring magnetic field strength, as a potential alternative to superconducting quantum interference devices (SQUIDs).
The measurement resolution is directly set by the radius of the hole in the TI.
This is of special interest because it provides a potential route for high-resolution magnetic field measurements even at room temperatures.\cite{PhysRevLett.111.136804, NatureMat.4384}

\section{Logic spintronics gates}
Next we note that the configuration in the previous section can also be used as a spintronics switch, with voltage used to encode 0 and 1.
The two leads can be used as source and drain, while the magnetic field is used as the gate.
From Eq.~\eqref{Equation:Conductivity} it is clear that a magnetic field strength $B = \frac{\hbar n}{qr^2}$ corresponds to on and off states for $n$ even and odd, respectively, and we therefore define the magnetic switching quantum
\begin{align}
	\label{Equation:Switching_quantum}
	b = \frac{\hbar}{qr^2}.
\end{align}

An alternative way to encode 1 and 0 is to use the currents of up- and down-spins, respectively.
This requires normal leads through which both up- and down-spins can be transported.
We therefore consider the same configuration, but now evaluate all four components of the transfer matrix $T(B, r, \pi/2)$:
\begin{align}
	T_{\uparrow\uparrow} = T_{\downarrow\downarrow} =& \cos\left(\frac{qBr^2\pi}{2\hbar}\right)e^{-ipr\pi/\hbar},\nonumber\\
	T_{\downarrow\uparrow} = -T_{\uparrow\downarrow} =& \sin\left(\frac{qBr^2\pi}{2\hbar}\right)e^{-ipr\pi/\hbar}.\nonumber
\end{align}

Similarly as above, the square of the transfer matrices gives the transfer probability of the spin-polarized currents.
In particular, the off-diagonal matrix elements $T_{\downarrow\uparrow} = -T_{\uparrow\downarrow}$ converts between up and down spin currents.
Therefore, the device relates the in and outgoing spin currents to each other through
\begin{align}
I_{\uparrow}^{out}		=& |T_{\uparrow\uparrow}|^2I_{\uparrow}^{in} + |T_{\uparrow\downarrow}|^2I_{\downarrow}^{in},\nonumber\\
I_{\downarrow}^{out}		=& |T_{\downarrow\uparrow}|^2I_{\uparrow}^{in} + |T_{\downarrow\downarrow}|^2I_{\downarrow}^{in}.\nonumber
\end{align}
Once again considering the special cases $B = \frac{\hbar n}{qr^2}$, with $n$ an integer, the currents transforms according to
\begin{align}
	I_{\uparrow}^{out}, I_{\downarrow}^{out}		=& \left\{\begin{array}{cc}
							I_{\uparrow}^{in}, I_{\downarrow}^{in}	& {\rm even}\; n,\\
							I_{\downarrow}^{in}, I_{\uparrow}^{in}	& {\rm odd}\; n.
								\end{array}\right.\nonumber
\end{align}
This means that the device can be switched between a normal lead and a NOT-gate, simply by changing $B$ by the switching quantum in Eq.~\eqref{Equation:Switching_quantum}.

\begin{figure*}
	\begin{center}
		\includegraphics[width=1\textwidth]{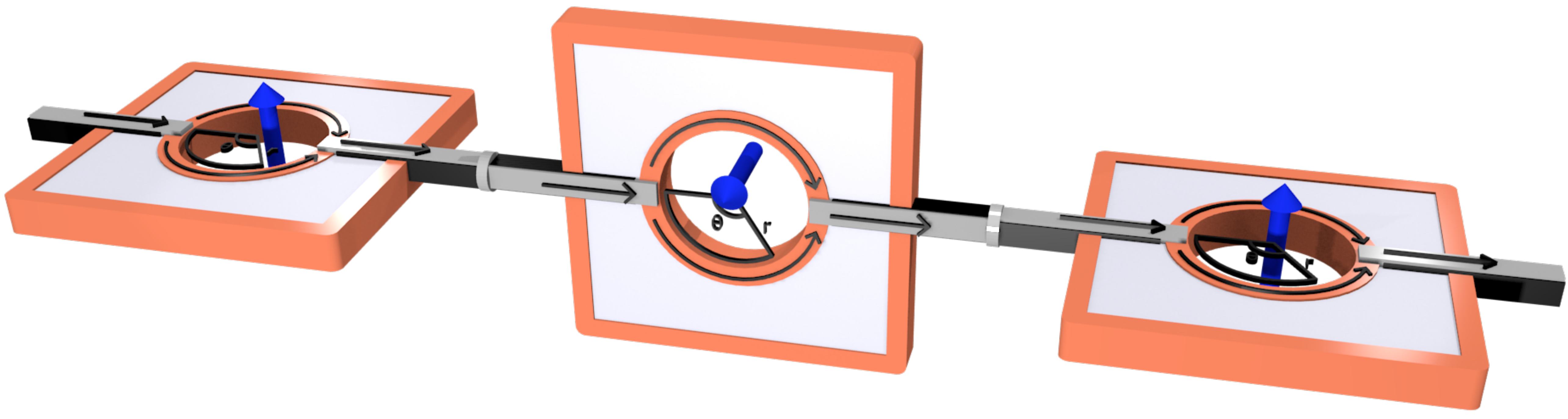}
	\end{center}
	\caption{Three solid state SG spin-splitters in series, with the middle device at an angle $\pi/2$ relative to the other two.}
	\label{Figure:Setup_QC}
\end{figure*}

\section{Quantum computer gate}
Having seen how an SG TI device can be used to construct classical logic gates for spintronics, we finally turn to possible applications in quantum computing.
It has been shown that a universal quantum computer can be built using only two-qubit CNOT-gates and single-qubit $SU(2)$-gates \cite{PhysRevA.52.5.52}.
We here show that a SG TI spin-splitter provides a route for implementing the later of these two gates.

For this purpose we consider three sequential spin-splitters connected by normal leads.
The three devices are oriented as in Fig.~\ref{Figure:Setup_QC}, with the middle device oriented at an angle $\varphi_2 = \pi/2$, while the first and last spin-splitters are at an angle $\varphi_1 = \varphi_3 = 0$.
The total transfer matrix for the complete system is then given by
\begin{align}
	T_{U(2)}	=& T(B_3, r_3, 0)T(B_2, r_2, \frac{\pi}{2})T(B_1, r_1, 0).\nonumber
\end{align}
When evaluated, this expression can be written as
\begin{widetext}
	\begin{align}
	\label{Equation:QC_Transfer_function}
		T_{U(2)} =& \left[\begin{array}{cc}
				e^{i\alpha}	& 0\\
				0			& e^{i\alpha}
			\end{array}\right]\left[\begin{array}{cc}
				e^{i\beta_{3}/2}	& 0\\
				0				& e^{-i\beta_{3}/2}
			\end{array}\right]\left[\begin{array}{cc}
				\cos(\frac{\beta_{2}}{2})	& \sin(\frac{\beta_{2}}{2})\\
				-\sin(\frac{\beta_{2}}{2})	& \cos(\frac{\beta_{2}}{2})
			\end{array}\right]\left[\begin{array}{cc}
				e^{i\beta_{1}/2}		& 0\\
				0				& e^{-i\beta_{1}/2}
			\end{array}\right],
	\end{align}
\end{widetext}
where $\alpha = -\frac{p(r_1 + r_2 + r_3)\pi}{\hbar}, \beta_i = -\frac{qB_ir_i^2\pi}{\hbar}$.

The six physical parameters $B_{i}, r_{i}$ are more than sufficient to make the four parameters $\alpha, \beta_{1}, \beta_{2}$ and $\beta_{3}$ independent of each other.
Moreover, when all these four parameters can be chosen independently, it is possible to express any $U(2)$-matrix using Eq.~(\ref{Equation:QC_Transfer_function}) \cite{PhysRevA.52.5.52}.
Thus, it is possible to implement any unitary single-qubit gate, and in particular any $SU(2)$-gate, through the use of three sequential solid state SG spin-splitters.
In fact, the overall $U(1)$-phase provided by the parameter $\alpha$ can be ignored for similar reasons that the $U(1)$-phase provided by the gauge transformation $A \rightarrow A + A'$ can be ignored.
This phase would only be relevant if the incoming electron is further split up into one part passing through the device, and one part moving through another path joining only at the far right outgoing lead.
In light of this it is useful to think of the devices as exhibiting a $SU(2)$-AB effect.
While the ordinary AB effect arise as a consequence of interference in a single $U(1)$-phase, these devices relies on a generalized $SU(2)$-interference effect in the relative phase and amplitude of the up and down components of the spin.
To be able to create an arbitrary $SU(2)$-transformation, a sequence of three devices is needed, while an individual spin-splitter gives rise to a subset of such $SU(2)$-transformations.
Finally, we note that in this calculation we have omitted transfer matrices describing the propagation through the leads. We are justified in doing so because these would be proportional to the identity and therefore only contribute to the irrelevant $\alpha$ phase.

\section{Discussion}
We would like to end with a few comments on some of the assumptions made.
First of all the tunnelling parameter $t$ which otherwise would have multiplied the $L$ and $R$ matrices was set to $t = 1$.
It is clear that the zeroth order correction to deviations from $t = 1$ is to include the factor $t^2$ in front of all transmission coefficients, which shows up as $t^4$ in the conductivity.
The higher order corrections would come from particles which are reflected and travels an additional time around the loop.
While such terms can introduce corrections to the interference pattern for intermediate field strengths, they would not affect the result at multiples of the switching quantum in Eq.~\eqref{Equation:Switching_quantum}.
The reason being that additional circuits around the loop will affect the relative phase between the up and down spins by multiples of $2\pi$.
Such interference effect could also play a role for $t = 1$ when ferromagnetic leads are used, because the down spins at the right edge will be completely reflected.
This can be solved by adding another ferromagnetic lead for the down spins to escape through at the right edge.

We also mention that although the setup in Fig.~\ref{Figure:Setup_QC} might seem prohibitively difficult to realize in practice, the focus has here been on providing a conceptually simple explanation of the phenomenon.
In fact, the only reason the middle spin-splitter is tilted at an angle $\pi/2$ is to make its edge states have their spin-polarization perpendicular to those of the other two.
In practice it would therefore be possible to have all three devices in the same plane, if it is constructed out of two different types of 2D TIs with perpendicular spin-polarization axes.

\section{Conclusion}
We have shown that the helical edge states of a 2D TI can be utilized to construct a solid state SG spin-splitter, which when threaded by a magnetic flux gives rise to a generalized $SU(2)$-AB interference effect.
With two ferromagnetic leads, the device can be used to accurately measure magnetic flux, as well as be used as a magnetic field gated spintronics switch.
Instead using normal leads, a switchable spintronics NOT-gate can be implemented, or when using three devices connected in sequence, a $SU(2)$-gate for quantum computing is achieved.

\section{Acknowledgements}
We are grateful to J.~Bardarson, J.~Fransson, and E.~Sj\"oqvist for useful discussions and the Swedish Research Council (Vetenskapsr\aa det), the G\"oran Gustafsson Foundation, and the Swedish Foundation for Strategic Research (SSF) for financial support.

\end{document}